\documentclass[9pt,twocolumn,twoside]{optica}
\setboolean{shortarticle}{true}
\setboolean{minireview}{false}
\usepackage{color}
\usepackage{amsmath}
\usepackage{graphicx}
\usepackage{multirow}
\usepackage{amsfonts}
\usepackage{amssymb}
\usepackage{amscd}
\usepackage{dcolumn}
\usepackage{bm}
\usepackage{float}
\usepackage{mathrsfs}
\usepackage{soul,color}
\soulregister\cite7
\soulregister\ref7

\newcommand{\ket}[1]{\mbox{$\left| #1 \right\rangle$}}

\title{Experimental nonlocal measurement of a product observable}

\author[1,2]{Yuan Li}
\author[1,2]{Han-Sen Zhong}
\author[1,2]{Yi-Han Luo}
\author[1,2]{Li-Chao Peng}
\author[1,2]{Chao-Yang Lu}
\author[1,2]{Nai-Le Liu}
\author[1,2]{Jun Zhang}
\author[1,2,*]{Li Li}
\author[1,2]{Jian-Wei Pan}

\affil[1]{Hefei National Laboratory for Physical Sciences at the Microscale and Department
of Modern Physics, University of Science and Technology of China, Hefei, Anhui 230026, China}
\affil[2]{CAS Center for Excellence Center in Quantum Information
and Quantum Physics, University of Science and Technology of China, Hefei, Anhui 230026, China}

\affil[*]{Corresponding author: eidos@ustc.edu.cn}

\ociscodes{(270.0270) Quantum optics.}

\begin{abstract}
Nonlocal measurement, or instantaneous measurement of nonlocal observables, is a considerably difficult task even for a simple form of product observable since relativistic causality prohibits interaction between spacelike separate subsystems. Following a recent proposal for effectively creating the von Neumann measurement Hamiltonian of nonlocal observables [Brodutch and Cohen, Phys. Rev. Lett. 116, 070404 (2016)], here we report a proof-of-principle demonstration of nonlocally measuring a product observable using linear optics without the violation of relativistic causality. Our scheme provides a feasible approach to perform nonlocal measurements via quantum erasure with linear optics.
\end{abstract}

\setboolean{displaycopyright}{true}

\begin{document}

\maketitle

\section{Introduction}
The compatibility of quantum mechanics and special relativity is a fundamental issue of physics. One of important connection points is the measurability of nonlocal observables. Quantum mechanics allows to perform nonlocal measurements, i.e., simultaneous measurements of nonlocal observables operated on spacelike separate subsystems, however, relativistic causality prohibits to perform such measurements. In 1931, Landau and Peierls~\cite{landau1931erweiterung} claimed the impossibility of measuring any nonlocal variable without the violation of relativistic causality. This conjecture was disproved later ~\cite{aharonov1981can,aharonov1986measurement,popescu1994causality}. Given certain system states, there exist a few types of observables that can be simultaneously measured under the restriction of relativistic causality. Then, approaches for nonlocal measurements have been studied, showing that all nonlocal observables can be measured instantaneously~\cite{groisman2002measurements,vaidman2003instantaneous,clark2010entanglement}.
However, these approaches rely on the verification measurements rather than the standard von Neumann measurements. A verification measurement can confirm whether the system is in an eigenstate of observable and produces the desired probabilities. Nevertheless, it does not necessarily leave the system in this eigenstate as the von Neumann measurement does, which indicates that these approaches are destructive and unrepeatable.

Measuring nonlocal product observables plays a significant role in quantum theory and quantum information processing, such as quantum nonlocality tests \cite{clauser1969proposed,bennett1999quantum}, semicausal measurements \cite{beckman2001causal}, error corrections \cite{gottesman1997stabilizer1}, and the interaction between two spins \cite{joulain2016quantum}.
Performing the standard von Neumann measurement of nonlocal observables is pretty difficult. Considering a simple product observable $A\otimes B$ on a bipartite system ${{\mathcal{H}}_{S}}={{\mathcal{H}}_{A}}\otimes {{\mathcal{H}}_{B}}$, where $A$ and $B$ are Hermitian operators on the Hilbert space ${{\mathcal{H}}_{A}}$ and ${{\mathcal{H}}_{B}}$, respectively, the standard von Neumann measurement requires an interaction Hamiltonian between the system and the meter.
This can be represented as $f\left( t \right)A\otimes B\otimes {{P}^{M}}$, where $f\left( t \right)$ is a function with a compact support near the time of measurement, $g=\int_0^\tau f(t)dt$ is the coupling strength, and ${{P}^{M}}$ is the conjugate momentum to the pointer variable of the meter. However, nonlocal measurements imply accessing spacelike separate subsystems, thus the von Neumann Hamiltonian modeling interaction between meters and nonlocal subsystems cannot be implemented directly.

Recently, Brodutch and Cohen proposed a protocol to effectively create the von Neumann measurement Hamiltonian for a large class of nonlocal observables~\cite{brodutch2016nonlocal}. With the help of an entangled ancillary state and the processing of quantum erasure~\cite{scully1982quantum,herzog1995complementarity,walborn2002double,peruzzo2012quantum}, the von Neumann measurement Hamiltonian $f\left( t \right)A\otimes B\otimes {{P}^{M}}$ can be created locally, which outputs the equivalent measurement results of desired probabilities and eigenstates of nonlocal observables
as von Neumann model, without violations of relativistic causality.

In this letter, we report a proof-of-principle implementation of the above protocol with linear optics, i.e., the nonlocal measurement of a product Pauli observable ${{\sigma }_{z}}\otimes {{\sigma }_{z}}$.
By adjusting the coupling strength between the system and the meter, in principle our method can also implement nonlocal weak measurement of product observables ${{\sigma }_{z}}\otimes {{\sigma }_{z}}$. Further, our method can be similarly applied for the nonlocal measurements of other product Pauli observables in linear optical system.

\section{The nonlocal measurement scheme}
\begin{figure}[htbp]%
	\centering
	\includegraphics[width=7cm]{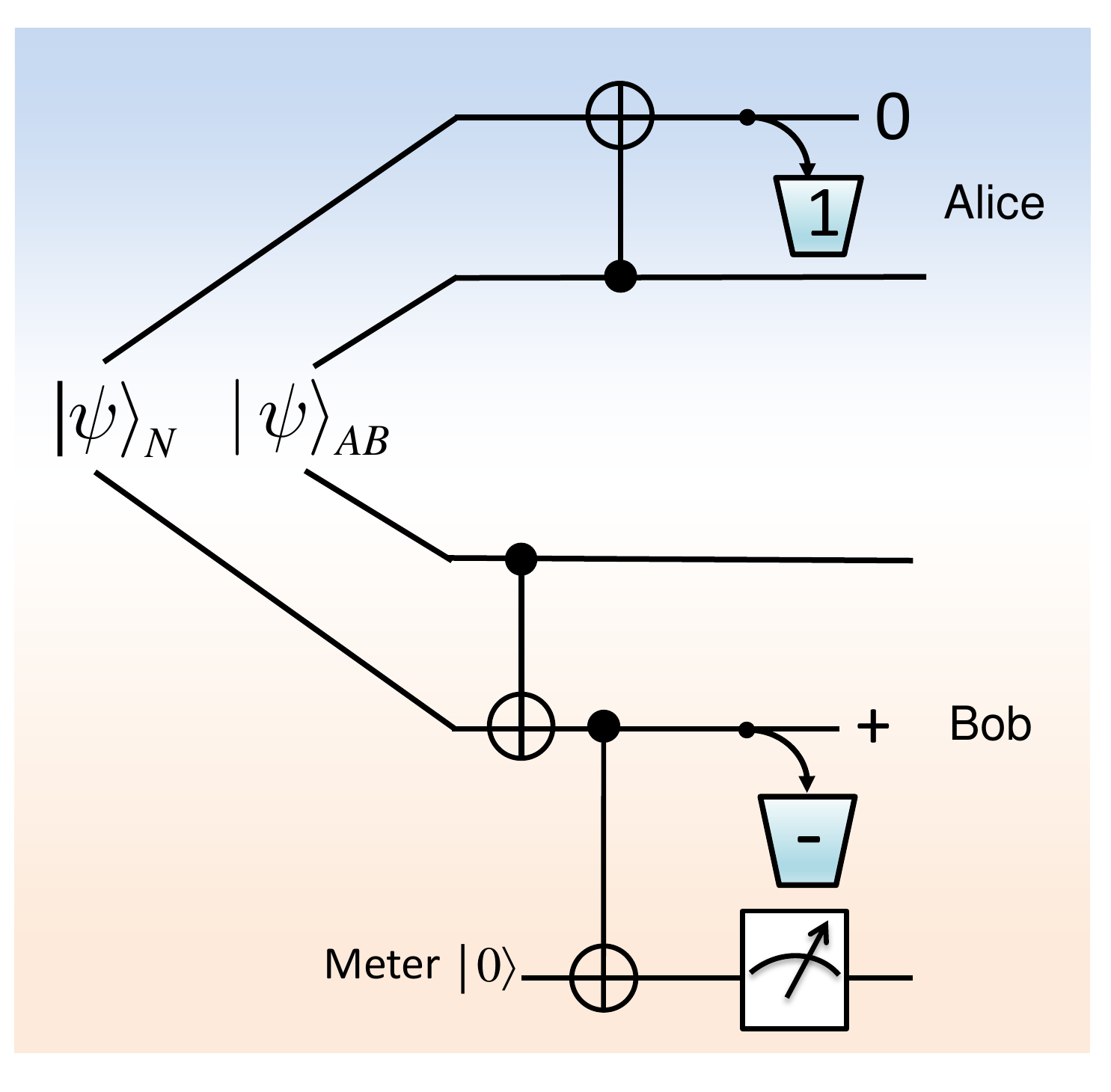}
	\caption{Scheme for nonlocal measurement of product Pauli observable $\sigma_{z} \otimes \sigma_{z}$. ${\left| \psi \rangle \right.}_{AB}$ and ${\left| \psi \rangle  \right.}_{N}$ represent the system state to be measured and an ancillary entangled state, respectively.
		By using three CNOT gates operated on the system qubits, the ancillary qubits and the local meter qubit,
		Alice and Bob retain only the measurement results of ${\left| 0 \rangle \right.}$ and ${\left| + \rangle  \right.}$, respectively, on the ancillary entangled state. Corresponding to the measurement result of local meter qubit, the output state of the observable $\sigma_{z} \otimes \sigma_{z}$ is obtained.}
	\label{fig1}
\end{figure}

The goal of scheme is to measure an observable operator ${{\sigma }_{z}}\otimes {{\sigma }_{z}}$ on an arbitrary pure state of two-qubit system, i.e., $ {{\left| \psi \rangle  \right.}_{AB}}={{a}_{1}}\left| 0 \right.{{\rangle }_{A}}{{\left| 0\rangle  \right.}_{B}}+{{a}_{2}}\left| 0 \right.{{\rangle }_{A}}\left| 1 \right.{{\rangle }_{B}}+{{a}_{3}}{{\left| 1\rangle  \right.}_{A}}{{\left| 0\rangle  \right.}_{B}}+{{a}_{4}}{{\left| 1\rangle  \right.}_{A}}\left| 1 \right.{{\rangle }_{B}}$.
Due to the degeneracy of its eignstates, the measurement output projects the system state either to the subspace ${{\left| \psi \rangle  \right.}_{+}}={{ \Pi }_{+}}\left| \psi  \right.{{\rangle }_{AB}}=1/\sqrt{|a_1|^2+|a_4|^2}({{a}_{1}}\left| 0 \right.{{\rangle }_{A}}\left| 0 \right.{{\rangle }_{B}}+{{a}_{4}}{{\left| 1\rangle  \right.}_{A}}\left| 1 \right.{{\rangle }_{B}})$ or to the subspace $\left| \psi  \right.{{\rangle }_{-}}={{ \Pi }_{-}}{{\left| \psi \rangle  \right.}_{AB}}=1/\sqrt{|a_2|^2+|a_3|^2}({{a}_{2}}{{\left| 0\rangle  \right.}_{A}}{{\left| 1\rangle  \right.}_{B}}+{{a}_{3}}{{\left| 1\rangle  \right.}_{A}}\left| 0 \right.{{\rangle }_{B}})$, corresponding to an eigenvalue of +1 or -1 with a probability of $|a_1|^2+|a_4|^2$ or $|a_2|^2+|a_3|^2$. Here,
$\Pi_{+}=| 00  \rangle \langle {00} |+|11\rangle \langle {11}|$ and $\Pi_{-}=\left| 01 \right.\left. \rangle \langle 01 \right|+\left| 10\rangle \right.\langle \left. 10 \right|$ are the Kraus operators for the measurement.
We note that both the measurement output states, i.e., $|\psi\rangle_+$ and $|\psi\rangle_-$, are entangled, which cannot be realized by local separate measurements of the observable $\sigma_z$ operated on subsystems $A$ and $B$ individually.

As proposed in Ref.~\cite{brodutch2016nonlocal}, the nonlocal measurement of the observable operator ${{\sigma}_{z}}\otimes {{\sigma}_{z}}$ on arbitrary pure state of two-qubit system $|\psi\rangle_{AB}$  can be realized by using an ancillary entangled state and an ancillary meter qubit.
The ancillary entangled state ${{\left| \psi  \right.\rangle}_{N}}=1/\sqrt{2}\left( {{\left| 0\rangle  \right.}_{{{N}_{A}}}}\left| 0 \right.{{\rangle }_{{{N}_{B}}}}+\left| 1 \right.{{\rangle }_{{{N}_{A}}}}\left| 1 \right.{{\rangle }_{{{N}_{B}}}} \right)$ is shared in prior between $A$ and $B$, and the ancillary meter qubit $\left| \psi  \right.{{\rangle }_{M}}={{\left| 0\rangle  \right.}_{M}}$ is located at $B$ to register the measurement results, as illustrated in Fig.~\ref{fig1}.
The process of nonlocal measurement scheme is listed as follows.

Step 1: Perform a controlled NOT (CNOT) gate operation between control qubit $A$ and target qubit ${{N}_{A}}$.

Step 2: Perform a measurement on qubit ${{N}_{A}}$ in $\{\ket{0}, \ket{1}\}$ basis and extract only the result of $\ket{0}$.

Step 3: Perform a CNOT gate operation between control qubit $B$ and target qubit ${{N}_{B}}$.

Step 4: Perform a CNOT gate operation between control qubit ${{N}_{B}}$ and target qubit $M$. After this step, the composite state between the system and the meter is
${{\left| \psi \rangle  \right.}_{4}}={{a}_{1}}\left| 0 \right.{{\rangle }_{A}}{{\left| 0\rangle  \right.}_{B}}\left| 0 \right.{{\rangle }_{{{N}_{B}}}}\left| 0 \right.{{\rangle }_{M}}+{{a}_{2}}\left| 0 \right.{{\rangle }_{A}}\left| 1 \right.{{\rangle }_{B}}{{\left| 1\rangle  \right.}_{{{N}_{B}}}}\left| 1 \right.{{\rangle }_{M}}+{{a}_{3}}\left| 1 \right.{{\rangle }_{A}}\left| 0 \right.{{\rangle }_{B}}{{\left| 1\rangle  \right.}_{{{N}_{B}}}}\left| 1 \right.{{\rangle }_{M}}+{{a}_{4}}{{\left| 1\rangle  \right.}_{A}}{{\left| 1\rangle  \right.}_{B}}{{\left| 0\rangle  \right.}_{{{N}_{B}}}}\left| 0 \right.{{\rangle }_{M}}$.

Step 5: Quantum erasure operation on qubit ${{N}_{B}}$ in $\{\ket{+}, \ket{-}\}$ basis ($\left| \pm  \right.\rangle =\left( \left| 0 \right.\rangle \pm \left| 1 \right.\rangle  \right)/\sqrt{2}$). For instance, given the projection measurement to the state $|+\rangle$, the composite state between the system and the meter is changed to
$\left| \psi  \right.{{\rangle }_{5}}=\left( {{a}_{1}}\left| 0 \right.{{\rangle }_{A}}\left| 0 \right.{{\rangle }_{B}}+{{a}_{4}}{{\left| 1\rangle  \right.}_{A}}\left| 1 \right.{{\rangle }_{B}} \right){{\left| 0\rangle  \right.}_{M}}\pm \left( {{a}_{2}}{{\left| 0\rangle  \right.}_{A}}\left| 1 \right.{{\rangle }_{B}}+{{a}_{3}}\left| 1 \right.{{\rangle }_{A}}\left| 0 \right.{{\rangle }_{B}} \right){{\left| 1\rangle  \right.}_{M}}$.

Step 6: Read out the state of meter qubit. The output results of $\ket{0}$ and $\ket{1}$ indicate that
the two-qubit system is projected to the subspaces corresponding to the operators $\Pi_{+}$ and $\Pi_{-}$, respectively.

The two cascaded CNOT operations performed in step 3 and step 4 are equivalent to the interaction $e^{-i\frac{\pi}{4}\sigma_z^B \sigma_z^{N_B} \sigma_x^{M}} e^{i \frac{\pi}{4}\sigma_z^B \sigma_z^{N_B}} e^{i  \frac{\pi}{4}\sigma_x^{M}}$ operating on qubits $B$, $N_B$ and $M$, in which the term  $e^{-i\frac{\pi}{4}\sigma_z^B \sigma_z^{N_B} \sigma_x^M}$ is exactly the coupling unitary evolution performed locally by Bob as required in the Brodutch-Cohen protocol. In addition, local single-qubit operation $e^{i  \frac{\pi}{4}\sigma_x^M}$ just changes the computational basis of the meter, and the term $e^{i \frac{\pi}{4}\sigma_z^B \sigma_z^{N_B}}$ compensates the additional phase $e^{i\frac{\pi}{2}}$ of the meter, to satisfy the convention in the protocol, i.e., $e^{\pm{i} \frac{\pi}{4} {P}^M} |q=\frac{\pi}{2}\rangle=|q=\frac{\pi}{2}\pm\frac{\pi}{4}\times2\rangle$, where ${P}^M=\sigma_x^M$ and $q$ represents the zenith angle of the meter qubit in the Bloch sphere. According to the protocol, after the quantum erasure step the nonlocal interaction $e^{-i\frac{\pi}{4}\sigma_z^A \sigma_z^{B} \sigma_x^M}$ operating on qubits $A$, $B$ and $M$ is equivalently performed, with a result of $|\psi\rangle_5$. Moreover, two controlled operations in step 3 and step 4 could be convenient to adjust the coupling strength $g$ from the strong domain to the weak domain. For instance, if one CNOT operation, i.e.,
controlled-$\sigma_x$ gate, is changed to controlled-$e^{-i\frac{\phi}{2}\sigma_x}$ gate, after
the quantum erasure step the interaction between $|\psi\rangle_{AB}$ and the meter
would be $e^{-i\frac{\phi}{4}\sigma_z^A \sigma_z^{B} \sigma_x^M}e^{i  \frac{\phi}{4}\sigma_x^M}$, where $e^{-i\frac{\phi}{4}\sigma_z^A \sigma_z^{B} \sigma_x^M}$ is the von Neumann measurement interaction and $e^{i  \frac{\phi}{4}\sigma_x^M}$ is the local unitary operation of the meter.

\section{Experiment}

\begin{figure}[htbp]
	\centering
	\includegraphics[width=9 cm]{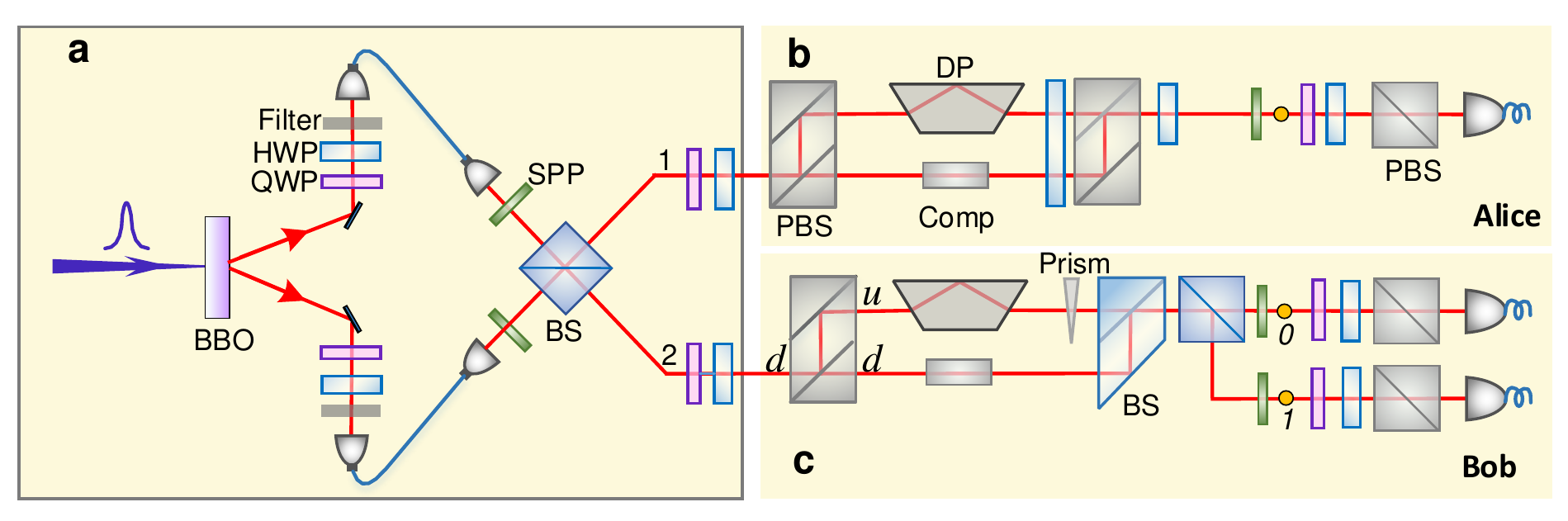}
	\caption{Experimental setup for nonlocal measurement. (a) Initial state preparation. An ultrafast laser beam with a central wavelength of 394 nm is focused on a BBO crystal to create photon pairs at 788 nm. Two SPPs and one BS are used to postselect two OAM entangled photons. One HWP and one QWP are placed in sequence at Alice and Bob, respectively, to prepare the initial polarization states. (b) Realization of the Polar-OAM CNOT gate (polarization mode is used as control qubit and OAM mode is used as target qubit) at Alice. (c) Realization of both the Polar-Path CNOT gate (polarization mode is used as control qubit and path mode is used as target qubit) and the Polar-OAM CNOT gate at Bob. The measurement of spatial qubit at $(|u\rangle+|d\rangle)/\sqrt{2}$ is performed with a prism and a BS whilst the OAM measurement is performed with SPPs. The tomography of polarization qubit is then performed with a set of QWP, HWP and PBS both at Alice and Bob. BBO: beta barium borate, SPP: spiral phase plate, BS: beam splitter, PBS: polarization beam splitter, HWP: half-wave plate, QWP: quarter-wave plate, DP: Dove prism, Comp: optical path compensation.}
	\label{fig2}
\end{figure}

In order to realize a proof-of-principle demonstration of nonlocal measurement with linear optics, we exploit the multiple degrees of freedom of photons~\cite{wang201818}, i.e., the polarization, the path, and the orbital angular momentum (OAM), where the polarization modes are used as system qubits and the other modes are used as auxiliary qubits.

The experimental setup is shown in Fig. \ref{fig2}.
In the experiment, photon pairs are generated simultaneously via the process of type-II spontaneous parametric down-conversion.
The down-converted photons are filtered by two 3 nm filters and coupled into single-mode fibers. After filtering by the single-mode fibers, the photon pairs are then collimated into the free space to prepare Gaussian OAM mode.
A combination of half-wave plate (HWP) and quarter-wave plate (QWP) are used to compensate the polarization of photons and thus to guarantee that
the photons before entering the spiral phase plate (SPP) in both sides are $H$-polarized.
The SPPs are used to transform the Gaussian OAM modes into right-handed OAM modes of $+\hbar$ (denoted as $|r\rangle$).

The photons in both sides arrive at a 50:50 beam splitter (BS) simultaneously to create perfect interference. The BS is polarization-preserving, and OAM-preserving in the transmission path and OAM-swapping in the reflection path between $|r\rangle$ and $|l\rangle$ (left-handed OAM modes of $-\hbar$). By postselection
the coincidence counts between the two output ports of BS indicate that a two-photon state $|H\rangle_1|H\rangle_2(|r\rangle_1|r\rangle_2-|l\rangle_1|l\rangle_2)$ is generated. A HWP and a QWP are placed at each side to regulate the polarization degree of freedom. Then,
the initial state is prepared as $|\psi\rangle_0=(a_1|HH\rangle+ a_2|HV\rangle+ a_3|VH\rangle+ a_4|VV\rangle)\otimes(|rr\rangle-|ll\rangle) \otimes |d\rangle$, where $|d\rangle$ is the inital state of meter qubit encoded at the down path mode.

\begin{figure}[tbp]%
	\centering
	\includegraphics[width=9 cm]{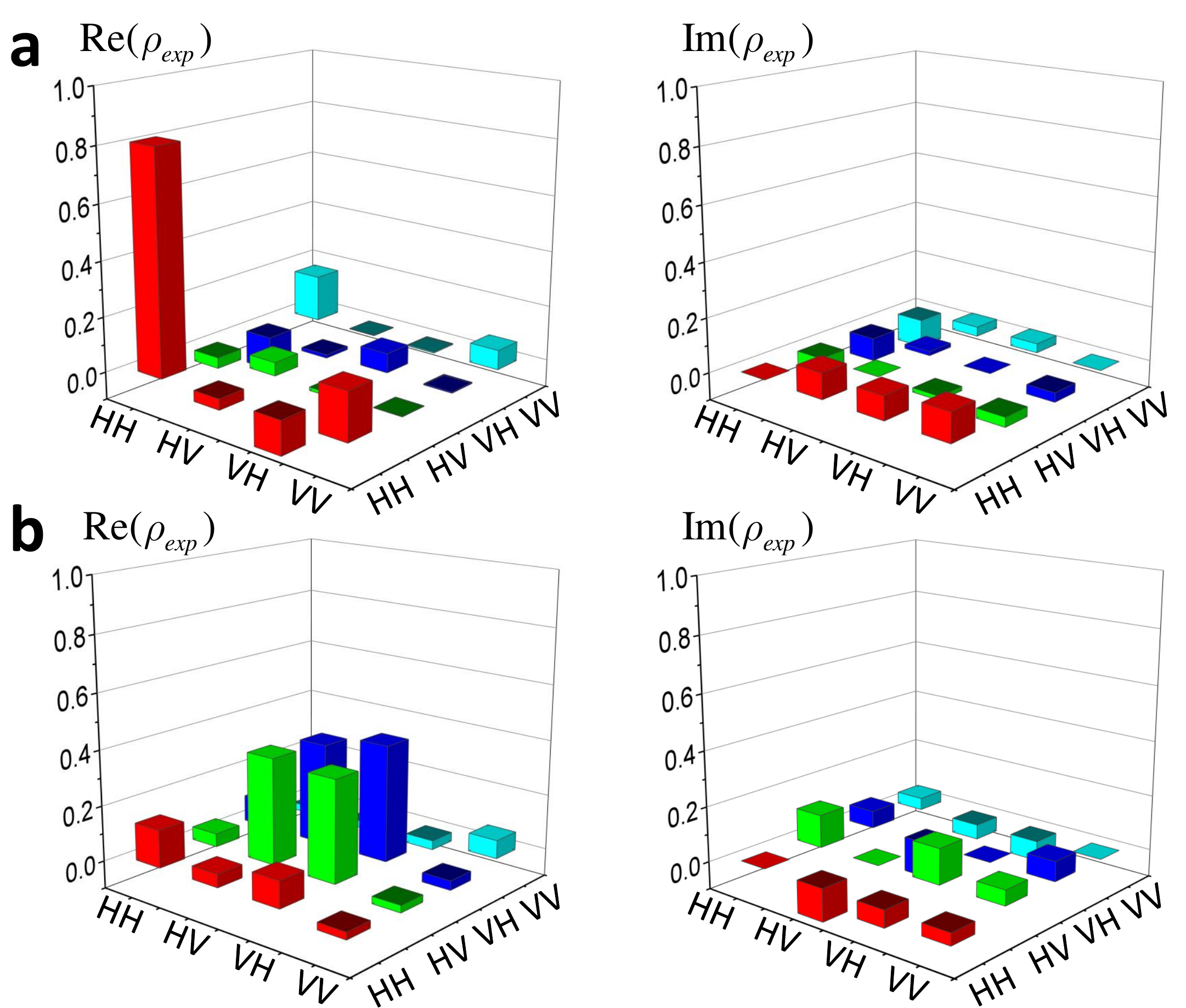}  %
	\caption{Reconstructed output density matrix for input state $|\phi\rangle_4$ including real part (left) and imaginary part (right), with the
		projection to the subspace of $\Pi_+$ (a) and to the subspace of $\Pi_-$ (b), respectively.}  %
	\label{fig3}
\end{figure}

The initial state is then sent to two parties, i.e., Alice and Bob as depicted in Fig. \ref{fig2}(b) and Fig. \ref{fig2}(c). At Alice's part,
a polarization beam splitter (PBS) is used to separate $H$ and $V$ polarized photons, and a Dove prism is inserted in the reflection path ($V$ polarization) to realize a NOT gate for the OAM qubit. An optical path compensation is inserted in the transmission path ($H$ polarization) to guarantee that photons in two paths arrive at the second PBS simultaneously.
In such a way, a CNOT gate is performed after the recombination of polarization modes by inserting two HWPs at $45^\circ$  ahead and behind the second PBS, in which polarization and OAM modes are operated as control and target qubits, respectively.
Further, a SPP is used to effectively read out the OAM measurement results by transforming $|r\rangle$ mode into $|G\rangle$ mode, where $|G\rangle$ denotes the
fundamental Gaussian mode that can be efficiently coupled into single-mode fiber.

At Bob's part, considering the fact that it is more convenient to treat the degree of freedom (DoF) of OAM as a target qubit rather than a control qubit for CNOT operation, the notations of qubit $N_B$ and qubit $M$ are swapped in the experiment, which does not affect the effectiveness of our scheme.
The incident photons with a polarization state of $|\varphi\rangle_B=\alpha|H\rangle+\beta|V\rangle$ pass through a PBS that separates the photons into the down path mode ($|d\rangle$) and up path mode ($|u\rangle$).
This can be regarded as a CNOT gate, in which polarization mode is operated as control qubit whilst path mode is operated as target qubit. Then the output state between the two degrees of freedom can be written as $\alpha|H\rangle|d\rangle+\beta|V\rangle|u\rangle$.

\begin{table*}[t]
	\centering
	\caption{The measurement results of fidelity and probability projected to the subspace of $\Pi_+$ and  $\Pi_-$ in the cases of four input states. The error bars of fidelity and probability are calculated by Poissionian counting statistics. }
	{\begin{tabular}{ccccccc}
			\toprule
			Input state &  State projected to $\Pi_+$ subspace & Fidelity &Probability& State projected to $\Pi_-$ subspace & Fidelity  &Probability\\
			\midrule
			$|\phi\rangle_1$	&	$|HH\rangle$&	0.878(5)&0.532(9)	&	$|HV\rangle$&	0.891(5) &0.467(9)\\
			$|\phi {{\rangle }_{2}}$	&	$(|HH\rangle +|VV\rangle )/\sqrt{2}$&	0.826(7)&0.517(11)&	$(|HV\rangle +|VH\rangle )/\sqrt{2}$.&	0.837(7) &0.483(11)\\
			$|\phi {{\rangle }_{3}}$	&	$(|HH\rangle +i|VV\rangle )/\sqrt{2}$&	0.829(7)&0.508(11)	&	$(i|HV\rangle +|VH\rangle )/\sqrt{2}$&	0.877(7)&0.492(11)\\
			$|\phi {{\rangle }_{4}}$	&	$(2|HH\rangle +|VV\rangle )/\sqrt{5}$&	0.801(8)&0.609(9)	&	$(|HV\rangle +|VH\rangle )/\sqrt{2}$&	0.740(9)&0.391(9)\\
			
			\bottomrule
	\end{tabular}}
	\label{table1}
\end{table*}

Given that the initial state at Alice's part is $(|H\rangle+|V\rangle)/\sqrt{2}$, after Alice's measurement of OAM mode, the two-photon composite state is changed to
$\alpha|H\rangle|Hdr\rangle+\alpha|H\rangle|Vdl\rangle+\beta|V\rangle|Hur\rangle+\beta|V\rangle|Vul\rangle$, from which one can find out that
the polarization mode of Alice and Bob is mapped into the path and OAM modes at Bob.
In order to further project the system state to the subspaces of ${{   \Pi   }_{+}}=\left| HH\rangle \langle  \right.\left. HH \right|+\left| VV \right.\rangle \langle \left. VV \right|$ or ${{   \Pi   }_{-}}=\left| HV\rangle \langle  \right.\left. HV \right|+\left| VH \right.\rangle \langle \left. VH \right|$, the measurements on path and OAM modes are performed at Bob's part via
another CNOT gate, where the path and OAM modes are used as control and target qubits, respectively.
A Dove prism is inserted in the up path to realize NOT gate of OAM mode, and a small-angle prism in the Mach-Zehnder interferometer is used to
realize the path mode measurement in the basis of $(|d\rangle+|u\rangle)/\sqrt{2}$. Then, one BS and two SPPs are used to measure the OAM modes of $|r\rangle$ and $|l\rangle$, respectively, which corresponds to the measurement results of operators $\Pi_+$ and $\Pi_-$. Finally, via the two-photon coincidence counts at Alice and Bob, a full tomography of two-photon state is performed
for the polarization mode with a combination of a QWP, a HWP and a PBS in each output port~\cite{agnew2011tomography}.

In the experiment, we choose four input states, i.e., $|\phi\rangle_1 =|+\rangle|H\rangle$,$|\phi\rangle_2=|+\rangle|+\rangle$,$|\phi\rangle_3=|+\rangle|R\rangle$ and $|\phi\rangle_4= (\sqrt{2}|H\rangle+|V\rangle)(\sqrt{2}|H\rangle+|V\rangle)/3$, to verify the nonlocal measurement, where $|R\rangle=(|H\rangle+i |V\rangle)/\sqrt{2}$. Fig. \ref{fig3} presents the measured real part and imaginary part of density matrix $\rho_{exp}$ in the case of $|\phi\rangle_4$, with the projection to the subspaces of $\Pi_+$ and ${{\Pi }_{-}}$.
Table~\ref{table1} lists the measurement results of fidelity and probability for four input state.
The fidelity of the output state is calculated with the definition of $F=Tr(\rho_{exp}\rho_{ideal})$.
The probability of the output state is calculated according to the coincidence counts in the $\{ \ket{H}, \ket{V} \}$ basis.

Compared with encoding single photon with only one DoF, using multiple DoFs simultaneously for single photon can significantly decrease the number of photons as required in the scheme, and can realize deterministic CNOT gate rather than probabilistic CNOT gate with single DoF~\cite{o2003demonstration,okamoto2005demonstration}. In addition, the preparation method of polarization-OAM entanglement in the experiment can be widely used for applications such as quantum purification~\cite{pan2001entanglement,li2010deterministic} and quantum cryptography~\cite{chen2006deterministic}.

\section{Conclusion}
In summary, we have experimentally demonstrated nonlocal measurement for a product Pauli observable ${{\sigma }_{z}}\otimes {{\sigma }_{z}}$. Our scheme can be easily adapted to perform other product Pauli observables by local unitary rotations. For instance, ${{\sigma }_{x}}$ can be treated as ${{\sigma }_{z}}$ operation in the basis of $\{ \ket{+}, \ket{-} \}$. Thus, to obtain the nonlocal measurement of ${{\sigma }_{x}}\otimes {{\sigma }_{z}}$ one can perform ${{\sigma }_{x}}$ operation on subsystem $A$ before step 1 and after step 6 as described in the schme.
Further, applying arbitrary unitary operation to OAM with recently developed technology \cite{wang201818} instead of NOT operation to OAM with a Dove prism as implemented in the experiment, coupling strength $g$ can be adjusted, so that nonlocal weak measurement of product Pauli observables can also be performed.
Our work presents the feasibility of nonlocal measurements of product observables with linear optics, which may have many potential applications in quantum foundation and quantum information processing.

\section{Funding Information}
National Key R$\&$D Program of China (2017YFA0304004); National Natural Science Foundation of China (91336214, 11574297, and 11374287); Chinese Academy of Sciences.
\section{Acknowledgement}
The authors acknowledge insightful discussions with X.-L.
Wang. Note added: recently, we became aware that a similar work has been implemented with different experimental setup~\cite{pan2019photonic}.

\end{document}